# AIR POLLUTION MODELLING USING A GRAPHICS PROCESSING UNIT WITH CUDA

F. Molnár Jr.[1], T. Szakály[1], R. Mészáros[1], I. Lagzi[1,2*]

[1] Department of Meteorology, Eötvös Loránd University, P.O. Box 32, H-1518 Budapest, Hungary

[2] Department of Chemical and Biological Engineering, Northwestern University, Evanston, Illinois, USA

**Abstract**

The Graphics Processing Unit (GPU) is a powerful tool for parallel computing. In the past years the performance and capabilities of GPUs have increased, and the Compute Unified Device Architecture (CUDA) – a parallel computing architecture – has been developed by NVIDIA to utilize this performance in general purpose computations. Here we show for the first time a possible application of GPU for environmental studies serving as a basement for decision making strategies. A stochastic Lagrangian particle model has been developed on CUDA to estimate the transport and the transformation of the radionuclides from a single point source during an accidental release. Our results show that parallel implementation achieves typical acceleration values in the order of 80–120 times compared to CPU using a single-threaded implementation on a 2.33 GHz desktop computer. Only very small differences have been found between the results obtained from GPU and CPU simulations, which are comparable with the effect of stochastic transport phenomena in atmosphere. The relatively high speedup with no additional costs to maintain this parallel architecture could result in a wide usage of GPU for diversified environmental applications in the near future.

---
[*] Corresponding author. Tel.: +36-2090555; fax: +36-1372-2904.
*E-mail address*: lagzi@vuk.chem.elte.hu (I. Lagzi).

**1. Introduction**

Simulation of the dispersion of chemically active or passive species (e.g. radionuclides) from a single and strong source is a great challenge for both computational and decision making strategy point of views. Model simulations must have a high degree of accuracy and must be achieved faster than real time to be of use in decision support [1]. Development of successful and cost efficient strategies requires a very accurate prediction of the dispersion of plume and concentration of species. Underestimation of the maximum dosage may have serious health consequences. On the other hand, applying remediation measures in regions where significant dosage will not be received would waste valuable resources and may have significant social implications if evacuation is required. Therefore, it is very important to develop methods incorporating appropriate modelling strategies, which are able to satisfy these criteria concerning computational time and error of numerical simulations [1].

Dispersion of chemical species from a single source can be calculated with different dynamic models. The predominant model types are the so-called Eulerian and Lagrangian transport models. Eulerian models have the advantage of being computed on a three-dimensional grid providing 3D descriptions of the meteorological fields rather than trajectories of single particles. Here the partial differential equations describing the variation of the concentrations of species over time can be solved with various numerical methods (e.g. Finite Element Method, Method of Lines, etc) [1–4]. The basis of the Method of Lines is the spatial discretisation of partial differential equations followed by time integration of produced ordinary differential equations with appropriate initial and boundary conditions. When traditionally used with fixed meshes, Eulerian models have difficulties in resolving steep gradients. This phenomenon causes problems particularly in resolving dispersion from a single point source, resulting in very large gradients near the release. However, there are several ideas to reduce this error using a nested grid with finer resolution only around the point source or application of adaptive grid [1,5,6]. Here algorithm can automatically place a finer resolution grid, adaptively in time and space, in the regions where higher spatial numerical error can be expected.

In contrast to the Eulerian approach, in Largangian models, the air masses or individual particles with assigned mass or Gaussian shaped puffs of pollutants travel along trajectories determined mainly by the wind field. Lagrangian models have the advantage of allowing high spatial resolution. Their potential disadvantages occur when long-range transport simulations result in strongly diverging flows leading to uncertainties in long-range trajectories. However, in this study, the dispersion of radionuclides was simulated on local scale, therefore the Lagrangian method is an adequate tool to solve this problem. In the model, the simulations of advection, diffusion and source and sinks of the substances require a huge computational effort.

There are numerous solutions to address the issue of high computational requirements. Generally, the solution is using supercomputers, clusters, or grid systems [7–16]. These systems are built by connecting numerous processors, either by some sort of direct link or by a network connection. Several computing centers can be connected to each other by the internet, thus creating grids. Management and programming of these systems require certain software environments and tools [17,18].

In the past five years, the technological development of consumer graphics hardware created the possibility to use desktop video cards to solve numerically intensive problems

[19–33], since their computational capacity far exceeds the capacity of desktop CPUs. Using GPUs (processors of video cards) for general purpose calculations is called GPGPU. Its main advantage is high cost-effectiveness. Compared to former solutions, there are virtually no extra costs to operate these systems, like high-powered air conditioning, high electric power consumption and the need for a large space (building). Programming GPU is achieved by using library functions designed for 3D computer graphics (games, CAD). Due to this constraint programming such applications is difficult. Recently, NVIDIA addressed this problem by creating a new parallel computing model called CUDA [29]. It is also an extension to the well-known C language. A compiler, a software development kit with utilities and numerous examples, and a complete documentation of the architecture and the language extensions are available [34].

In this paper we present an efficient parallelization of the stochastic Lagrangian particle model using CUDA. In this model, each particle can be handled independently, thus being a perfect candidate for parallelization.

## 2. Model description

Stochastic Lagrangian particle model was chosen and developed for simulating accidental release on local scale because this model can handle high gradient of species near the point source as discussed in the Introduction part. In this model type, individual particles are released from a point source in each time step. Each particle represents a given mass or activity, and they can be moved in 3D space by advection (wind field) and turbulent diffusion, they can transform to other species (by radioactive decay), and particles can be removed from the atmosphere via dry and wet deposition processes to the surface. This means that the model is continuous in space and discontinuous in time. It was supposed that advection field is deterministic, but the effect of the turbulent diffusion is always stochastic. The new spatial position of particles can be calculated with the following equations [35]:

$$X_i^{new} = X_i^{old} + v_i^{adv}\Delta t + \tilde{x}_i, \tag{1}$$

where $X_i^{new}$ and $X_i^{old}$ are the spatial coordinates of particles after and before the time step, respectively. $v_i^{adv}$ is the i$^{th}$ coordinate of the wind velocity vector, $\Delta t$ is the time step (in the model simulation 10 s). The last term in Eq. (1) ($\tilde{x}_i$) describes the effect of stochastic turbulent processes, and $i$ denotes the spatial dimension ($i$ = 1, 2, 3). Stochastic term is calculated by

$$\tilde{x}_i = (rand)\sqrt{2K_i\Delta t}. \tag{2}$$

Random numbers (*rand*) with normal distribution (with mean 0.0 and variance 1.0) were generated using Mersenne Twister random number generator [36] and a Cartesian Box-Muller transformation [37]. $K_i$ is the turbulent diffusion coefficient in each direction. For horizontal dispersion a constant value was used ($K_x = K_y = 100\ \text{m}^2\text{s}^{-1}$). Vertical diffusion coefficient ($K_z$) depends on the height, and it was calculated by the following expression:

$$K_z(z) = \frac{ku_*z}{\Phi\left(\frac{z}{L}\right)}\left(1 - \frac{z}{H_z}\right)^2. \tag{3}$$

Here $k$, $u_*$, $H_z$, $\Phi\left(\frac{z}{L}\right)$ and $L$ are von Kármán's constant, friction velocity, mixing layer height, similarity function for heat and Monin–Obukhov length, respectively and $z$ is the height of interest. Friction velocity and Monin–Obukhov length were calculated iteratively in the function of the actual vertical stratification of the atmosphere [38].

Moreover, released radioactive particles can decay and deposit. Radioactive decay was deterministic in the model based on half-time of radionuclides. Deposition processes of particles whilst they can remove from the atmosphere to the surface by dry and wet pathways were assumed to be stochastic. During dry deposition, particles can deposit below the mixing layer height. Wet deposition only occurs in case of precipitation. The detailed description of these processes is out of scope of this paper, and will be discussed elsewhere.

## 3. Parallel application

### 3.1. Concepts of CUDA

The main concept of CUDA parallel computing model is to operate with tens of thousands of lightweight *threads*, grouped into *thread blocks*. These threads must execute the same function, with different parameters. The function that contains the computations and runs in parallel in many instances is called the *kernel*. Threads in the same thread group can synchronize with each other, by inserting synchronization points in the kernel, which must be reached by all threads in the group before continuing execution. These threads can also share data during execution. This way usually several hundred threads in the same block can work cooperatively. Threads of different thread blocks cannot be synchronized and should be considered to run independently.

It is possible to use a small number of threads and/or small number of blocks to execute a kernel, however, it would be very inefficient. This would utilize only a fraction of the computing power of the GPU. Therefore, CUDA is the best suited to those problems that can be divided into many parts, which can be computed independently (in different blocks), and these should be further divided into smaller cooperating pieces (into threads).

There are several types of memory available in CUDA, designed for different uses in kernels. Proper use of these memories can increase the computation performance. The *global memory* is essentially the random access video memory available on the video card. It may be read or written any time at any location by any of the threads, but to achieve high performance access to global memory should be coalesced, meaning the threads must follow a specific memory access pattern. For the complete (and hardware-revision dependent) description please refer to the Programming Guide [39]. A kernel has access to two cached, read-only memories: the *constant memory* and the *texture memory*. Constant memory may be used to store constants that do not change during kernel execution. Texture memory may be used efficiently when threads access data with spatial locality in one, two, or three dimensions. It also provides built-in linear interpolation of the data. There is also a parallel data cache available for reading and writing for all the threads of a thread block called the *shared memory*. It makes the cooperative work of threads in a block possible. It is divided into banks. Kernels should be written in a way to avoid bank conflicts, meaning the threads which are executed physically at the same time should access different banks. The Programming Guide presents details how to achieve this [39].

Memory management and kernel execution is controlled by CUDA library functions in the *host* code (the one which runs on the CPU). While the kernels are executing on the *device*, the CPU continues to execute host code, so CPU and GPU can work in parallel.

*3.2 Application*

Implementing a single-threaded CPU version of the model is straightforward. Assuming that the emission profile (the amount of emitted particles in every time step) for all species and the maximum number of particles released during the simulation are known, activities or masses can be a priori assigned to the particles. The maximum number of particles is the dominant variable that affects simulation time and precision. It is limited only by available memory.

The main loop is the time evolution of the simulation (Figure 1). In every iteration the main steps for every particle are the same: interpolating (sampling) weather data in $x$, $y$, $z$, and time dimensions using linear interpolation, calculating the turbulent diffusion coefficient, moving the particle by the wind, then by turbulent diffusion, and finally testing for dry and wet depositions. Particles may become inactive, meaning they are no longer moved by wind or turbulent diffusion, when the particle is deposited or it reaches the predefined boundaries of the simulated area. At the end of every $n^{th}$ time step the activity of particles is calculated based on the isotope's half-life and the time since particle was emitted. Activities or masses of particles are summarized on a rectangular grid for visualization and further statistical evaluation. From the technical point of view, the value of $n$ should not be smaller than four, it should be set to the highest number possible to achieve the desired precision of the time-integrated dosage calculation (a post-processing of simulation results).

In the parallelized CUDA version of the program, we utilize the various memory types available. Weather data are loaded into three-dimensional textures, to utilize the hardware-implemented trilinear interpolation. Since the weather data are four-dimensional, an extra interpolation step is necessary. The fourth interpolation must be in $z$ dimension, because the vertical weather information is nonlinear, unlike $x$, $y$ and time dimensions. Using texture memory is also useful because the plume usually propagates in a specific direction, giving the required spatial locality for the texture cache to work efficiently. Physical constants of the isotopes are loaded to constant memory. Shared memory is used for caching the particle data (position, state information). The data is loaded from global memory to shared memory when a kernel starts, and written back at the end, if data was modified.

The calculations in each time step are done by two kernel functions. First step is generating random numbers, using the Mersenne-Twister random number generator [36]. The implementation is provided by the CUDA SDK. A large buffer (on the GPU) is used to store the numbers, because the random number generator works more efficiently if large amount of numbers is asked for at once. The second step is the main kernel. First, particle information and random numbers are loaded from global memory to shared memory. Second step is sampling the weather data, interpolation in $z$ dimension, and calculating the turbulent diffusion coefficient. These results are stored in shared memory. Next step is moving the particles by wind field and turbulent diffusion, and testing for deposition. This step uses the interpolated weather prediction model outputs and the random numbers for calculation of turbulent motion and stochastic deposition. The final step is writing the changed particle information back to global memory.

Calculation and integration of activities on a rectangular grid is performed by the same way as in the sequential version, and it is calculated on the CPU, while the GPU is processing the next time steps in parallel. It should be noted that this step requires extra memory transfer between device and host and should be done as rarely as possible.

The kernel is configured to execute in 32 blocks, and with 256 threads in each block. The optimal number of threads in a block was found using the CUDA Occupancy Calculator, which is part of the CUDA SDK. All threads process multiple particles, since the number of particles may well exceed the number of threads.

Source code for both CPU and GPU versions are freely available to download from a webpage [40], terms of use are included on this page.

**4. Results and discussion**

Generally, a well parallelized algorithm must produce exactly the same results as the sequential version. However, it is only true, if the parallel computing system is built by units of the same platform (e.g. x86) as the one used to run the non-parallelized calculation. A CUDA-compatible GPU is a different platform, it is only capable of single precision computation (except for the newest video cards, which provide double precision with low performance) and although it follows many points of the IEEE-754 floating-point standard, precision of the mathematical functions is usually even lower, as it is explicitly described in the Programming Guide [39]. We reveal that the numerical error arising from this inaccuracy, compared to a common CPU, is within tolerable limits, and the result of the simulation produced by the GPU is equally valid as the result of the CPU version.

For the tests, two sets of weather data were used. One is referred as "real" weather situation. Fields of meteorological data are obtained from the ALADIN limited area weather prediction model [41]. The other is referred as "artificial" weather, in which the meteorological fields are completely constant and uniform (constant wind, temperature, etc.). The emission is always supposed to be the same: only one type of isotope ($^{131}$I) is emitted with a constant emission rate. The actual emission rate is 1 GBq / 10 min, but this value is irrelevant for the tests, since the emission is always scaled to the given total number of particles. The time step is 10 seconds in these simulations. Figure 2 depicts the plume structure of the radioactive species (Fig. 2a) and the total activity at the surface layer (Fig. 2b) originated from the point source (Paks Nuclear Power Plant which is located at the center of Hungary: 46°34' N, 18°51' E) 6 hours after a hypothetical accident in case of real meteorological fields. The plume structure is predominantly determined by the wind field, however, other meteorological parameters can also affect plume structure (e.g. vertical temperature gradient, planetary boundary layer height, etc.).

### 4.1. Comparison of CPU and GPU results

For the comparison of results obtained from calculation on CPU and GPU, the actual distance between two given particles were determined. For this purpose, we modified the parallelized version to use the same set of random numbers in the same way as in the sequential version with the same initial random number seed. Using the same random seed a pair of simulations was performed with 2160 particles. For every given particle, the distance between the CPU and GPU result was calculated. We repeated this comparison one hundred times using in each pair of simulations with the same initial random number seed, and the distances for the 2160 particles were separately averaged. Figure 3 and 4 show the results of the comparison of simulations in case of the real and the artificial weather conditions. Particles are numbered as they are emitted, therefore the first particles fly for the longest time, and their final position varies mostly. The last particles are only moved a short distance after emission, therefore very small differences can be observed in their positions. It is conspicuous that the real weather causes much more deviations, although the same mathematical functions are used, giving the same numerical errors. In case of a real weather, the wind velocity vectors (among all other parameters) were different from position to position. A small numerical error in calculating the next position of a particle (arising from differences between the CPU and the GPU mathematical functions) involves that a different wind vector will be used in the next iteration step causing divergence of particle trajectories. More complex weather situation may cause more deviation between the results calculating on CPU and GPU (Fig. 3a). In our case, the real weather meteorological fields describe very variable weather, the main direction of the wind turned almost 180 degrees during six hours. In case of the homogeneous artificial weather, the difference of particle positions is minimal (Fig. 3b), since a small error in calculating the particle's new position does not involve usage of different wind vectors. On the time scale considered in this study, these deviations caused by numerical errors are in the order of length scale of the turbulent diffusion (several tens meters) which is random by itself.

The effect of deviation of particle positions on the final results of the simulations was also investigated, which were considered to be the activities summarized on a rectangular grid. We performed five pairs of simulations, each pairs using the same random number sequence. The number of particles was $2.16 \times 10^6$. Activities of the particles were summarized on a 3D grid of 128×128×64 cells, representing a 80×80×0.3 km space. Five CPU and GPU results were separately averaged for all grid cells. Figure 4a and 4b show the relative difference of the total activity at the first (surface) layer between CPU and GPU results in both weather situations. This layer has been chosen because of its importance on human health. Probability density functions of the relative differences in both cases indicate that ~90% of these differences are less than 2.5%. These are in the order of the effect of the stochasticity of the turbulent diffusion on particle position.

### 4.2. Performance

The simulation is clearly limited by memory bandwidth, because there are only relatively few arithmetic operations to update a particle information. Therefore, the most important rule to optimize this simulation in CUDA is to follow the specific memory access pattern to achieve *coalesced* memory reads and writes. High multiprocessor occupancy (number of threads which are executed physically in parallel, compared to the maximum supported number of threads) is also desired to hide memory latency. The highest possible occupancy is 33% because of the high register requirement of the kernel, which is achieved by using 256 threads per block. Shared memory bank conflicts never arise since particles never interact with each other.

We performed simulations with various particle numbers on a 2.33 GHz Core 2 Duo CPU and on two video cards: GeForce 8800 GTS and GeForce 8800 GTX. Figure 5 shows the actual speedup gained using the graphics processors. There is no technological difference between the two cards, only computing power and memory speed. The ratio of the speedup with the GTS and the GTX cards well represents the performance difference between them. In both cases the speedup is higher as the particle number is increased. For optimal performance at least half million particles are required. If fewer particles are used, the GPU spends most of the time on initialization for the calculations, synchronizing execution, etc., while the useful calculations actually take less time. In a real application, however, one should use as many particles are allowed by the memory on the video card, which should be several millions of particles.

The speedup achieved is a result of many factors. The memory bandwidth between GPU and video card's RAM is approximately one order of magnitude higher than between CPU and PC's RAM. During GPU calculation access to physical constants and meteorological data is even faster, because constant and texture cache are used. Calculation is also accelerated by the instant trilinear interpolation provided by the GPU. Finally, the random number generation on the GPU is about two orders of magnitude faster than using the CPU.

The performance was measured without summarizing activities on a grid. The required frequency of summarization should be determined by the end-user of the application based on our model implementation. More frequent summarization will result in more precise radioactive dose calculation, but it will show down the simulation significantly. Some parts of the current activity and dose calculations may be computed by the GPU, in a separate kernel, cutting down the tasks of the CPU. The summarization could also be optimized to use all cores of the multi-core CPU's which are very common in nowadays.

5. Conclusion

Simulation of either air pollution formation or accidental release is one of the most challenging computational tasks because of its numerical complexity and simulation time. The numerical simulations must be obviously achieved faster than in real time in order to use them in decision support. A feasible way is the parallelization of the source code. We provide here a new framework for air quality modelling using a Graphics Processing Unit as a parallel environment. In case of a stochastic Lagrangian particle model the typical speedup is around 80–120 depending on particle number used. Comparisons of CPU and GPU results emphasize that there are some differences due to different computational platforms. Higher differences between GPU and CPU results can be observed in case of using real and complex meteorological conditions than in case of homogeneous data fields. However, the monitored differences in the position of the individual particles between of GPU and CPU results (~ 50 m) are in the order of characteristic length scale of the smallest relevant transport phenomena (turbulent diffusion) in the atmosphere. This indicates that the Graphics Processing Unit would be a promising and cost efficient tool to run parallel applications for air quality management.

**Acknowledgement**

Authors acknowledge the financial support of the Hungarian Research Found (OTKA K68253). This work makes use of results produced by the SEE-GRID eInfrastructure for regional eScience, a project co-funded by the European Commission (under contract number 211338) through the Seventh Framework Program. SEE-GRID-SCI stimulates widespread eInfrastructure uptake by new user groups extending over the region of South Eastern Europe, fostering collaboration and providing advanced capabilities to more researchers, with an emphasis on strategic groups in seismology, meteorology and environmental protection. Full information is available at http://www.see-grid-sci.eu

**Figure captions**

**Figure 1**

Flowchart of the stochastic Lagrangian model on CPU and GPU.

**Figure 2**

(a): The three-dimensional structure of the plume in the domain of 40 km × 40 km × 135 m and (b): the time-integrated activity of the $^{131}$I in the surface level 6 hours after a hypothetical accident. The height of the point source is 20 m. Meteorological fields were obtained from the ALADIN numerical weather prediction model. The $^{131}$I was continuously emitting from the Paks NPP with the constant rate. The simulation period was 6 hours from 16:00 UTC, 03 October, 2008 to 22:00 UTC, 03 October, 2008.

**Figure 3**

Average distance of individual particles between the CPU and GPU results in 3D using the same initial random number seed with meteorological fields (a): obtained from the ALADIN numerical weather prediction model and (b): homogeneous and uniform fields with the following meteorological parameters: $T$ = 20.0 °C (temperature), $v_x$ = 5.0 m s$^{-1}$ (wind component at $x$ direction), $v_y$ = 0 m s$^{-1}$ (wind component at $y$ direction), $v_z$ = 0 m s$^{-1}$ (wind component at $z$ direction), $H_{mix}$ = 120 m (height of the mixing layer), $z_0$ = 0.25 m (roughness length), $RH$ = 70.0% (relative humidity), $N$ = 0.0 (cloudiness). The height of the point sources in both cases is 20 m.

**Figure 4**

Relative spatial difference of activities calculated by CPU and GPU, compared to the CPU results in case of (a): "real" (same as in Fig 3a) and (b): "artificial" (same as in Fig3b) weather. Uncertain results indicate that either CPU or GPU results have a confidence interval higher than ±80% of the activity value (with 95% confidence level) and it would be meaningless to compare them to other results.

**Figure 5**

The speedup of the applications as a function of released particles using GeForce 8800 GTS and GeForce 8800 GTX video cards.

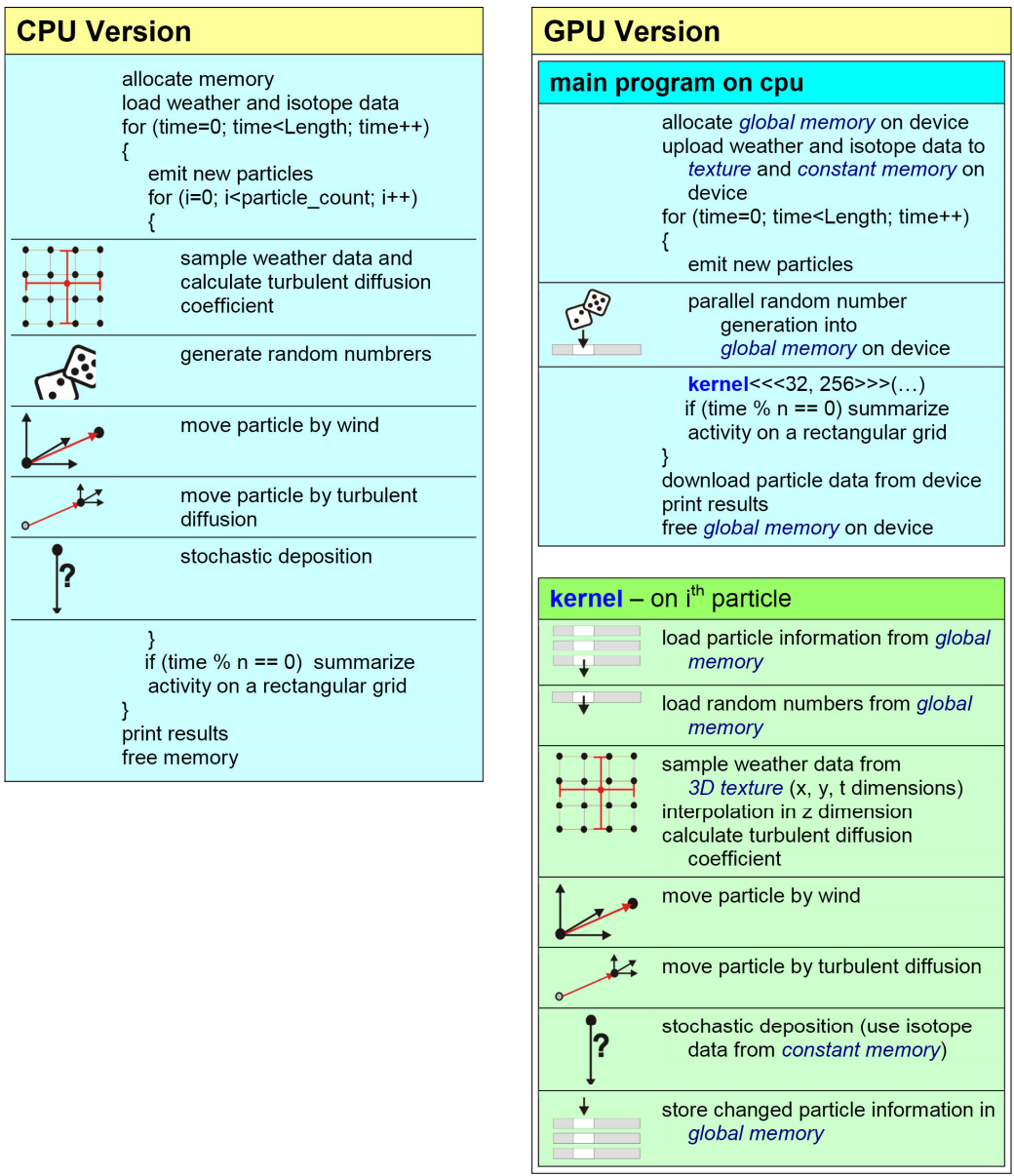

**Figure 1**

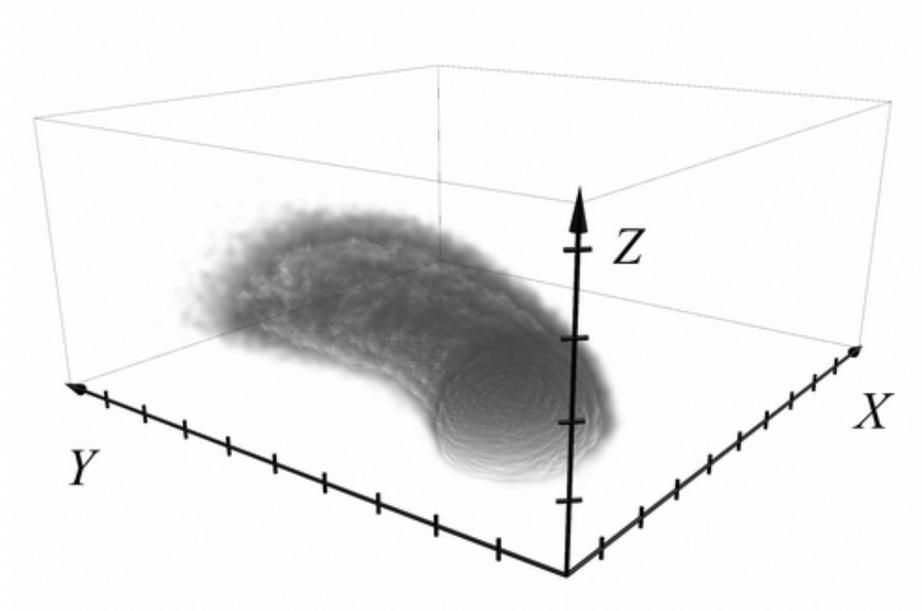

**Figure 2a**

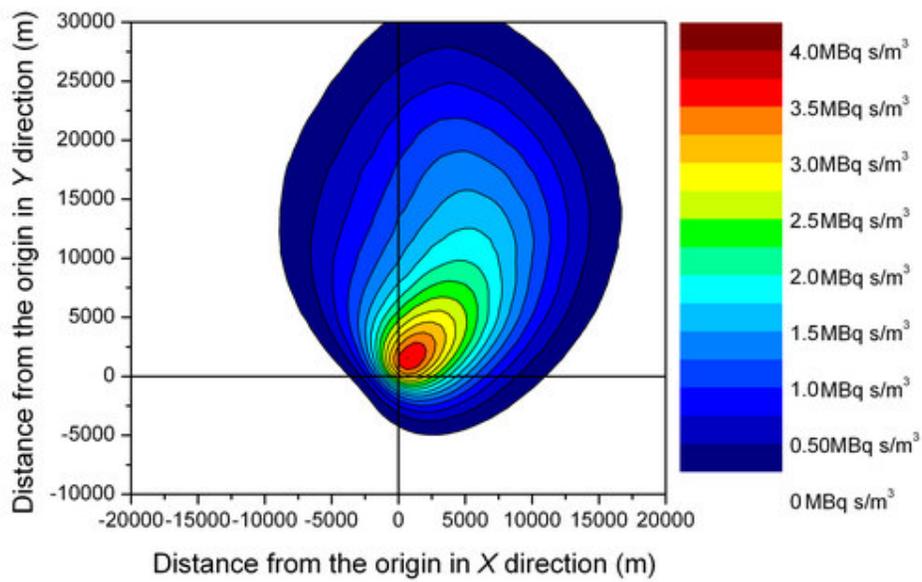

**Figure 2b**

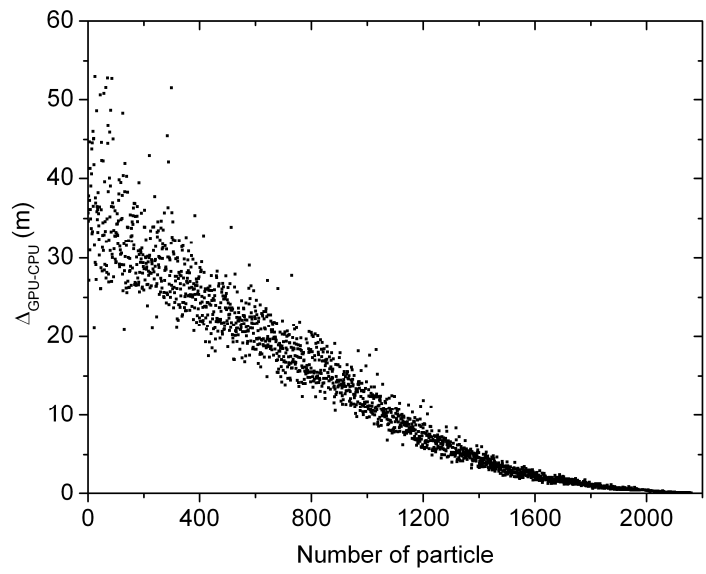

**Figure 3a**

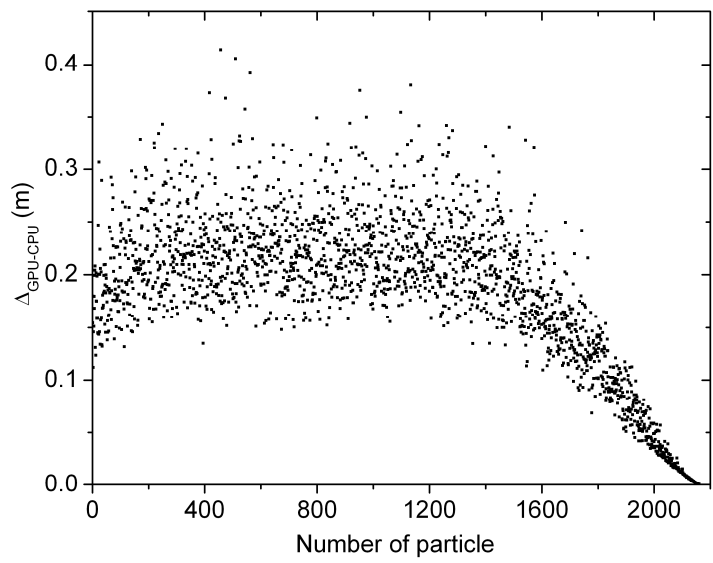

**Figure 3b**

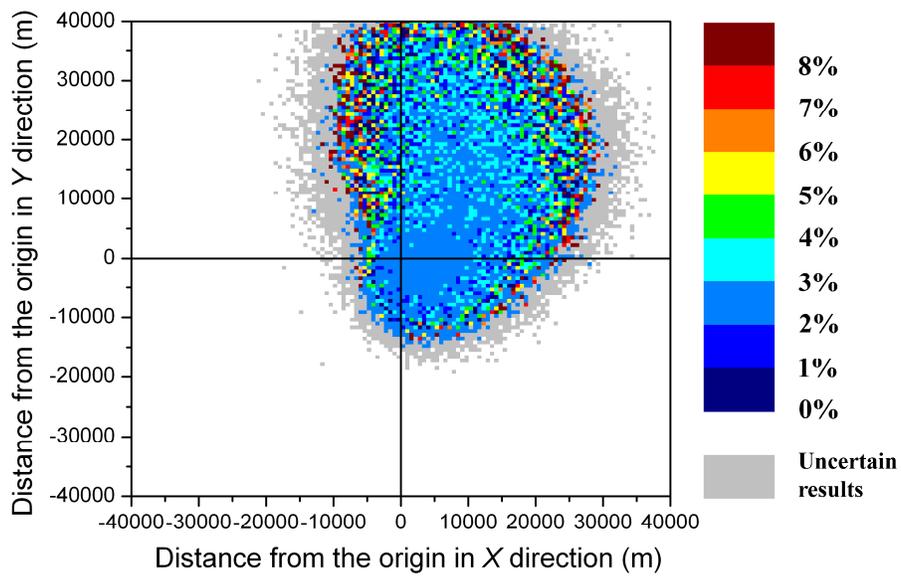

**Figure 4a**

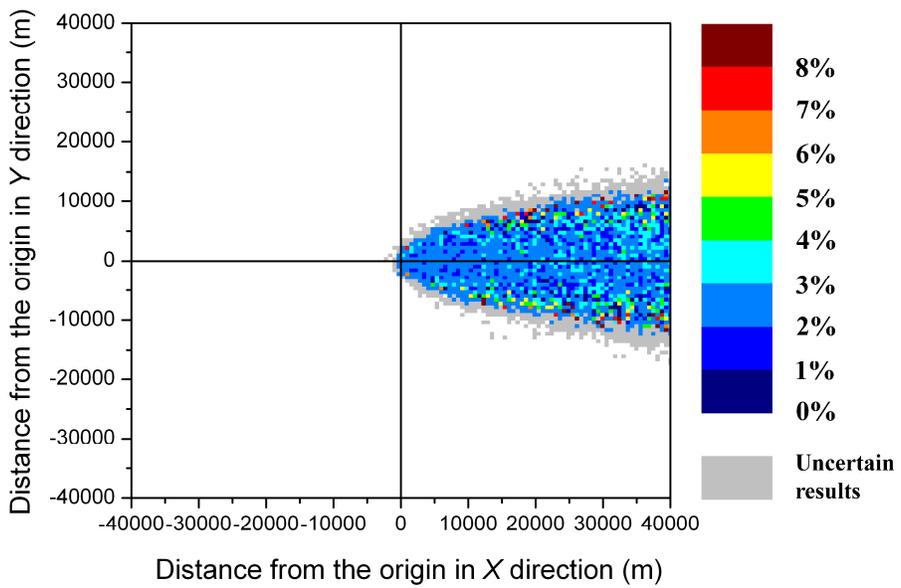

**Figure 4b**

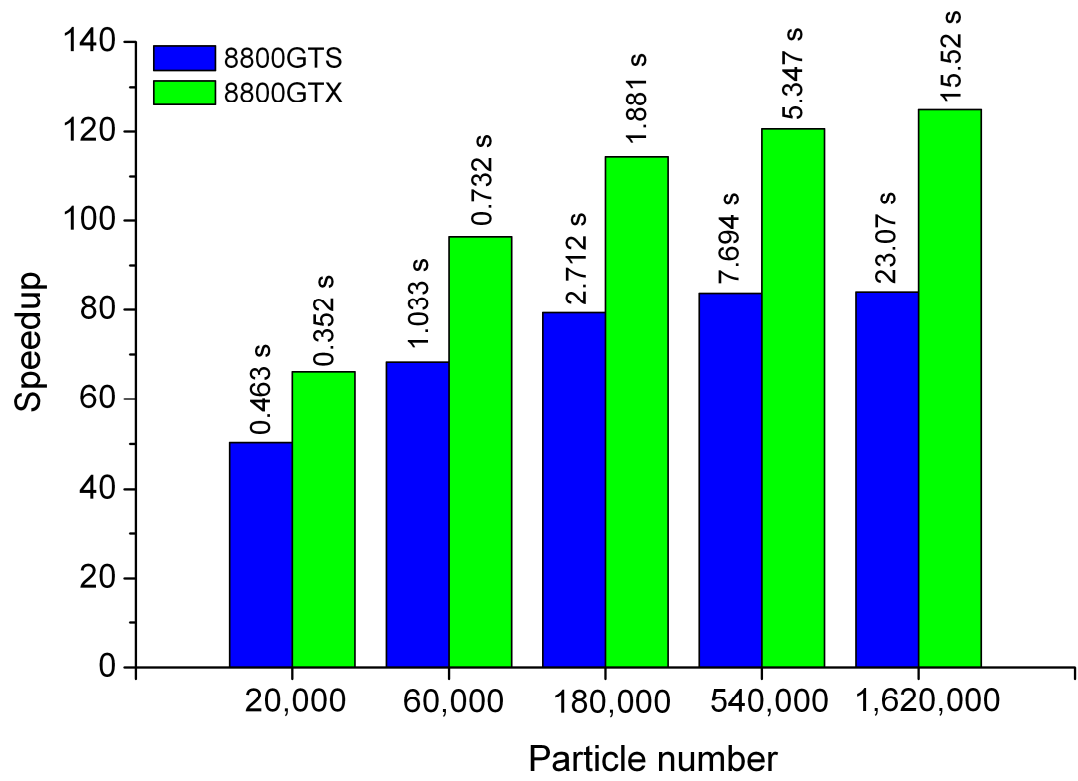

**Figure 5**